\documentclass[8pt,english]{article}
\newcommand{\be}{\begin{equation}}
\newcommand{\ee}{\end{equation}}
\newcommand{\ba}{\begin{eqnarray}}
\newcommand{\ea}{\end{eqnarray}}

\usepackage[english]{babel}
\usepackage{amsfonts,latexsym}
\usepackage{graphicx}
\usepackage{latexsym}
\usepackage[latin1]{inputenc}

\begin{document}


\def\halb{\frac{1}{2}}
\def\del2#1{\Delta_{#1,s}^{(2)}}

\title{Discrete Simulation of Power Law Noise}

\author{Neil Ashby\\
University of Colorado, Boulder CO\\ 
email:  ashby@boulder.nist.gov\\
Affiliate, NIST, Boulder, CO}
\maketitle

\begin{abstract}

	A method for simulating power law noise in clocks and oscillators is presented based on modification of the spectrum of white phase noise, then Fourier transforming to the time domain. Symmetric real matrices are introduced whose traces--the sums of their eigenvalues--are equal to the Allan variances, in overlapping or non-overlapping forms, as well as for the corresponding forms of the modified Allan variance.  Diagonalization of these matrices leads to expressions for the probability distributions for observing a variance at an arbitrary value of the sampling or averaging interval $\tau$, and hence for estimating confidence in the measurements.  A number of applications are presented for the common power-law noises.  
\end{abstract}

\section{Introduction}
The characterization of clock performance by means of average measures such as Allan variance, Hadamard variance, Theo, and modified forms of such variances, is widely applied within the time and frequency community as well as by most clock and oscillator fabricators.  Such variances are measured by comparing the times $t_k$ on a device under test, with the times at regular intervals $k\tau_0$ on a perfect reference, or at least on a better reference.  Imperfections in performance of the clock under test are studied by analyzing noise in the time deviation sequence $x_k=t_k-k\tau_0$, or the fractional frequency difference during the sampling interval $\tau=s\tau_0$:
\be
\Delta_{k,s}^{(1)}=(x_{k+s}-x_k)/(s\tau_0).
\ee
The frequency spectrum of fractional frequency differences can usually be adequately characterized by linear superposition of a small set of types of power law noise.  The frequency spectrum of the fractional frequency differences of a particular noise type is given by a one-sided spectral density 
\be\label{onesided}
S_y(f)=h_{\alpha}f^{\alpha},\quad f>0.
\ee
(The units of $S_y(f)$ are Hz$^{-1}$.)  For the common power-law noise types, $\alpha$ varies in integral steps from +2 down to -2 corresponding respectively to white phase modulation, flicker phase modulation, white frequency modulation, flicker frequency modulation, and random walk of frequency.   

	Simulation of clock noise can be extremely useful in testing software algorithms that use various windowing functions and Fourier transform algorithms to extract spectral density and stability information from measured time deviations, and especially in predicting the probability for observing a particular value of some clock stability variance.  This paper develops a simple simulation method for a time difference sequence that guarantees the spectral density will have some chosen average power law dependence.  Expressions for the common variances and their modified forms are derived here that agree with expressions found in the literature, with some exceptions.  This approach also leads to predictions of probabilities for observing a variance of a particular type at particular values of the sampling time.  A broad class of probability functions naturally arises.  These only rarely correspond to chi-squared distributions.

	This paper is organized as follows.  Sect. 2 introduces the basic simulation method, and Sect 3 applies the method to the overlapping Allan variance. Sect. 4 shows how diagonalization of the averaged squared second-difference operator, applied to the simulated time series, leads to expressions for the probability of observing a value of the variance for some chosen value of the sampling or averaging time.  Expressions for the mean squared deviation of the mean of the variance itself are derived in Sect. 6.  The approach is used to discuss the modified Allan variance in Sect. 7, and the non-overlapping form of the Allan variance is treated in Sect. 8.   Appendix 1 discusses evaluation of a contour integral for the probability, first introduced in Sect. 4, in the general case.

\section{Discrete Time Series}

	We imagine the noise amplitudes at Fourier frequencies $f_m$ are generated by a set of $N$ normally distributed random complex numbers $w_n$ having mean zero and variance $\sigma$, that would by themselves generate a simulated spectrum for white phase noise.  These random numbers are divided by a function of the frequency, $\vert f_m \vert^{\lambda}$, producing a spectral density that has the desired frequency characteristics.  For ordinary power law noise, the exponent $\lambda$ is a multiple of $1/2$, but it could be anything.   The frequency noise is then transformed to the time domain, producing a time series with the statistical properties of the selected power law noise.  The Allan variance, Modified Allan variance, Hadamard Variance, variances with dead time, and other quantities of interest can be calculated using either the frequency noise or the time series.  

	In the present paper we discuss applications to calculation of various versions of the Allan variance.  Of considerable interest are results for the probability of observing a value of the Allan variance for particular values of the sampling time $\tau$ and time series length $N$.  The derivations in this paper are theoretical predictions.  A natural frequency cutoff occurs at $f_h=1/(2 \tau_0)$, where $\tau_0$ is the time between successive time deviations.  This number is not necessarily related in an obvious way to some hardware bandwidth.  The measurements are assumed to be made at the times $k\tau_0$, and the time errors or residuals relative to the reference clock are denoted by $x_k$.  The averaging or sampling time is denoted by $\tau=s\tau_0$, where $s$ is an integer.  The total length of time of the entire measurement series is $T=N\tau_0$.  The possible frequencies that occur in the Fourier transform of the time residuals are
\be
f_m=\frac{m}{N\tau_0}\,.\quad\quad -\frac{N}{2}+1 \le m \le \frac{N}{2}\,.
\ee  

	{\it Noise Sequences}\/.  In order that a set of noise amplitudes in the frequency domain represent a real series in the time domain, the amplitudes must satisfy the reality condition
\be
w_{-m}=(w_m)^{\ast}.
\ee
$N$ random numbers are placed in $N/2$ real and $N/2$ imaginary parts of the positive and negative frequency spectrum.  Thus if $w_m=u_m+iv_m$ where $u_m$ and $v_m$ are independent uncorrelated random numbers, then $(w_m)^{\ast}=u_m-iv_m$.  Since the frequencies $\pm 1/2\tau_0$ represent essentially the same contribution, $v_{N/2}$ will not appear.  We shall assume the variance of the noise amplitudes is such that
\be\label{wmaverage}
\big<(w_m)^*w_n\big>=\big<u^2+v^2\big>\delta_{mn}=2\sigma^2 \delta_{mn};\quad m\ne 0,N/2.
\ee 
Also, $<w_m^2>=<u^2-v^2+2iuv>=0 {\rm\ for\ }m\ne N/2.$
The index $m$ runs from $-N/2+1$ to $N/2$.  In order to avoid division by zero, we shall always assume that the Fourier amplitude corresponding to zero frequency vanishes.  This only means that the average of the time residuals in the time series will be zero, and has no effect on any variance that involves time differences.

	We perform a discrete Fourier transform of the frequency noise and obtain the amplitude of the $k^{th}$ member of the time series for white PM:
\be
x_k=\frac{\tau_0^2}{\sqrt{N}}\sum_{m=-N/2+1}^{N/2} e^{-\frac{2\pi i m k}{N}}w_m.
\ee
The factor $\tau_0^2$ is inserted so that the time series will have the physical dimensions of time if $w_m$ has the dimensions of frequency.  We then multiply each frequency component by $\vert f_0/f_m \vert^{\lambda}$.  This will generate the desired power-law form of the spectral density.  The time series will be represented by
\be
X_k=\frac{\tau_0^2}{\sqrt{N}}\sum_{m=-N/2+1}^{N/2}\frac{\vert f_0 \vert^{\lambda}}{\vert f_m \vert^{\lambda}} e^{-\frac{2\pi i m k}{N}}w_m.
\ee
The constant factor $\vert f_0 \vert^{\lambda}$ has been inserted to maintain the physical units of the time series.  The noise level is determined by $f_0$.  For this to correspond to commonly used expressions for the one-sided spectral density, Eq. (\ref{onesided}), we shall assume that
\be
\frac{\tau_0^2\vert f_0 \vert^{\lambda}}{\sqrt{N}}=\sqrt{\frac{h_{\alpha}}{16 \pi^2 \sigma^2 (N \tau_0)}}\,.
\ee
We shall show that if $2 \lambda = 2-\alpha$ the correct average spectral density is obtained. The simulated time series is 
\be\label{timeseries}
X_k=\sqrt{\frac{h_{\alpha}}{16 \pi^2 \sigma^2 (N \tau_0)}}
\sum_m \frac{e^{-\frac{2\pi i m k}{N}}}{\vert f_m \vert^{\lambda}}w_m\,.
\ee
The average (two-sided) spectral density of the time residuals is obtained from a single term in Eq. (\ref{timeseries}):
\be
s_x(f_m)=\frac{h_{\alpha}}{16 \pi^2 \sigma^2 (N \tau_0) f_m^{2\lambda}}\bigg<\frac{w_mw_m^*}{\Delta f}\bigg>=\frac{h_{\alpha}}{8\pi^2 f_m^{2\lambda}}
\ee
where $\Delta f = 1/(N\tau_0)$ is the spacing between successive allowed frequencies.  The average (two-sided) spectral density of fractional frequency fluctuations is given by the well-known relation
\be
s_y(f)=(2 \pi f)^2 s_x(f)\,,
\ee
and the one-sided spectral density is

\be
S_y(f)=\cases{0, & $f < 0$;\cr
		2 s_y(f)= h_{\alpha}f^{\alpha},& $f>0$\,,\cr}
\ee
where $2 \lambda = 2 -\alpha$.
\section{Overlapping Allan Variance}
	Consider the second-difference operator defined by
\be
\Delta_{j,s}^{(2)}=\frac{1}{\sqrt{2 \tau^2}}(X_{j+2s}-2X_{j+s}+X_j).
\ee
The fully overlapping Allan variance is formed by averaging the square of this quantity over all possible values of $j$ from 1 to $N-2s$.  Thus
\be\label{overlapvar}
\sigma_y^2(\tau)=\bigg<\frac{1}{N-2s}\sum_{j=1}^{N-2s}(\Delta_{j,s}^{(2)})^2 \bigg>.
\ee
In terms of the time series, Eq. (\ref{timeseries}), the second difference can be reduced using elementary trigonometric identities:
\eject
\ba\label{del2ave}
\Delta_{j,s}^{(2)}=\sqrt{\frac{h_{\alpha}}{32 \pi^2 \tau^2 \sigma^2 (N\tau_0)}}\hbox to 2.3in{}\nonumber\\
\times \sum_m\frac{w_m}{\vert f_m \vert^{\lambda}}\bigg(e^{-\frac{2 \pi i m(j+2s)}{N}}-2e^{-\frac{2 \pi i m(j+s)}{N}}+e^{-\frac{2 \pi i m(j)}{N}}\bigg)\nonumber\\
=-\sqrt{\frac{h_{\alpha}}{2 \pi^2 \tau^2 \sigma^2 (N \tau_0)}}\sum_m\frac{w_m}{\vert f_m \vert^{\lambda}}e^{-\frac{2 \pi i m(j+s)}{N}}\bigg(\sin\frac{\pi m s}{N}\bigg)^2\,.
\ea
We form the averaged square of $\Delta_{j,s}^{(2)}$ by multiplying the real quantity times its complex conjugate, then averaging over all possible values of $j$.
\ba
\sigma_y^2(\tau)=\frac{h_{\alpha}}{2 \pi^2 \tau^2 (N\tau_0)(N-2s)}
\sum_{m,n,j}\bigg<\frac{w_m w_n^*}{\sigma^2}\bigg>
\frac{\bigg(\sin\big(\frac{\pi m s}{N}\big)\sin\big(\frac{\pi n s}{N}\big)\bigg)^2}{\vert f_m f_n \vert^{\lambda}}\nonumber\\
\hbox to -.5in{}\times e^{-\frac{2 \pi i (m-n)(j+s)}{N}}\,.
\ea
The average of the product of random variables only contributes $2 \sigma^2$ when $m=n$ (see Eq. (\ref{wmaverage}), except when $m=n=N/2$ where $\big<\vert w_{N/2} \vert^2\big>=\sigma^2$.
The Allan variance reduces to
\be
\sigma_y^2(\tau)=\frac{h_{\alpha}}{\pi^2 \tau^2(N\tau_0)}\bigg(\sum_m \frac{\bigg(\sin\big(\frac{\pi m s}{N}\big)\bigg)^4}{\vert f_m \vert^{2\lambda}} +\frac{1}{2}\frac{\bigg(\sin\big(\frac{\pi s}{2}\big)\bigg)^4}{\big(f_{N/2}\big)^{2\lambda}}\bigg)\,,
\ee
since every term in the sum over $j$ contributes the same amount.  The zero frequency term is excluded from the sum.  For convenience we introduce the abbreviation
\be
K=\frac{2 h_{\alpha}}{\pi^2 \tau^2 (N \tau_0)}\,. 
\ee
If we sum over positive frequencies only, a factor of 2 comes in except for the most positive frequency and so
\be
\sigma_y^2(\tau)=K\bigg(\sum_{m>0}^{N/2-1}\frac{\bigg(\sin\big(
\frac{\pi m s}{N}\big)\bigg)^4}
{f_m^{2\lambda}}+\frac{1}{4}\frac{\bigg(\sin\frac{\pi s}{2}\bigg)^4}{(f_{N/2})^{2\lambda}} \bigg).
\ee
If the frequencies are spaced densely enough to pass from the sum to an integral, then $\Delta f= (N \tau_0)^{-1}$ and
\be
\frac{1}{N\tau_0}\sum_mF(\vert f_m \vert)\rightarrow\int F(\vert f \vert)\,df
\ee
and we obtain the well-known result\cite{barnes71}
\be
\sigma_y^2(\tau)=2\int_0^{f_h}\frac{S_y(f)\,df}{\pi^2\tau^2f^2}\bigg(\sin(\pi f \tau)\bigg)^4\,.
\ee
Similar arguments lead to known expressions for the non-overlapping version of the Allan variance as well as for the modified Allan variance.  These will be discussed in later sections.
\section{Confidence Estimates}
	In the present section we shall develop expressions for the probability of observing a particular value $A_o$ for the overlapping Allan variance in a single measurement, or in a single simulation run.  $A_o$ is a random variable representing a possible value of the overlapping variance.  We use a subscript ``o" to denote the completely overlapping case.  To save writing, we introduce the following abbreviations:
\ba\label{FGdefs}
F_m^j=\frac{\bigg(\sin\bigg(\frac{\pi m s}{N}\bigg)\bigg)^2}{\vert f_m \vert^{\lambda}}\cos\bigg(\frac{2\pi m(j+s)}N{}\bigg)\nonumber\\
G_m^j=\frac{\bigg(\sin\bigg(\frac{\pi m s}{N}\bigg)\bigg)^2}{\vert f_m \vert^{\lambda}}\sin\bigg(\frac{2\pi m(j+s)}N{}\bigg)
\ea
The dependence on $s$ is suppressed, but is to be understood.  We write the second difference in terms of a sum over positive frequencies only, keeping in mind that the most positive and the most negative frequencies only contribute a single term since $\sin(\pi(j+s))=0$.  The imaginary contributions cancel, and we obtain
\be\label{diffFG}
\del2{j}=\sqrt{K}\sum_{m>0}\big(F_m^j \frac{u_m}{\sigma}+G_m^j\frac{v_m}{\sigma}\big)\,.
\ee
There is no term in $v_{N/2}$.  It is easy to see that the overlapping Allan variance is given by
\be\label{avarFG}
\sigma_y^2(\tau)=\frac{K}{N-2s}\sum_j\sum_{m>0}\bigg((F_m^j)^2+(G_m^j)^2 \bigg)\,.
\ee
To compute the probability that a particular value $A_o$ is observed for the Allan variance, given all the possible values that the random variables $u_1,v_1,...u_{N/2}$ can have, we form the integral
\be
\label{probdef}
P(A_o)=\int\delta\bigg(A_o-\frac{1}{N-2s}\sum_j \big(\Delta_{j,s}^{(2)}\big)^2\bigg)\prod_{m>0}\bigg(e^{-\frac{u_m^2+v_m^2}{2\sigma^2}}\frac{du_m dv_m}{2\pi \sigma^2}\bigg)\,.
\ee
The delta function constrains the averaged second difference to the specific value $A_o$ while the random variables $u_1,v_1,...u_m,v_m,...u_{N/2}$ range over their (normally distributed) values. There is no integral for $v_{N/2}$.  Inspecting this probability and the Eq. (\ref{diffFG})for the second difference indicates that we can dispense with the factors of $\sigma^{-1}$ and work with normally distributed random variables having variance unity.  Henceforth we set $\sigma=1$.

The exponent involving the random variables is a quadratic form that can be written in matrix form by introducing the $N-1$ dimension column vector $U$ (the zero frequency component is excluded)
\be
U^T=[u_1\,v_1\,...u_m\,v_m,...v_{N/2-1},u_{N/2}]\,.
\ee
Then
\be
\halb \sum_{m>0}(u_m^2+v_m^2)=\halb U^TU=\halb U^T {\bf 1} U,
\ee
where ${\bf 1}$ represents the unit matrix.  The delta-function in Eq. (\ref{probdef}) can be written in exponential form by introducing one of its well-known representations, an integral over all angular frequencies $\omega$:\cite{lighthill}
\be
P(A_o)=\int_{-\infty}^{\infty} \frac{d \omega}{2 \pi}e^{i\omega\big(A_o-\frac{1}{N-2s}\sum_j \big(\Delta_{j,s}^{(2)}\big)^2 \big)}\prod_{m>0}\bigg(e^{-\frac{u_m^2+v_m^2}{2\sigma^2}}\frac{du_m dv_m}{2\pi \sigma^2}\bigg)\,.
\ee
The contour of integration goes along the real axis in the complex $\omega$ plane.  

	The squared second difference is a complicated quadratic form in the random variables $u_1,v_1,...u_m,v_m,...u_{N/2}$.  If this quadratic form could be diagonalized without materially changing the other quadratic terms in the exponent, then the integrals could be performed in spite of the imaginary factor $i$ in the exponent.  To accomplish this we introduce a column vector $C^j$ that depends on $j,m,s,N$ and whose transpose is 
\ba
(C^j)^T=[F_1^j,G_1^j,...F_m^j,G_m^j,...G_{N/2-1}^j,F_{N/2}^j]\,.
\ea
Dependence on $s$ is not written explicitly but is understood.  The column vector has $N-1$ real components.  It contains all the dependence of the second difference on frequency and on the particular power law noise. We use indices $\{m,n\}$ as matrix (frequency) indices.   The (scalar) second difference operator can be written very compactly as a matrix product
\be
\del2{j}=\sqrt{K} (C^j)^TU=\sqrt{K}U^TC^j.
\ee
Then 
\be
\frac{1}{N-2s}\sum_j\bigg(\Delta_{j,s}^{(2)} \bigg)^2=U^T\bigg(\frac{K}{N-2s}\sum_j C^j(C^j)^T\bigg)U.
\ee
The matrix 
\be\label{hoverlapping}
H_o=\frac{K}{N-2s}\sum_jC^j(C^j)^T
\ee
is real and symmetric.  $H_o$ is also Hermitian and therefore has real eigenvalues.  A real symmetric matrix can be diagonalized by an orthogonal transformation,\cite{Strang,Stoll} which we denote by $O$.  Although we shall not need to determine this orthogonal transformation explicitly, it could be found by first finding the eigenvalues $\epsilon$ and eigenvectors $\psi_{\epsilon}$ of $H_o$, by solving the equation 
\be
H_o\psi_{\epsilon}=\epsilon \psi_{\epsilon}\,.
\ee
The transformation $O$ is a matrix of dimension $(N-1)\times (N-1)$ consisting of the components of the normalized eigenvectors placed in columns.  Then
\be
H_oO=OE\,,
\ee
where $E$ is a diagonal matrix with entries equal to the eigenvalues of the matrix $H_o$.  Then since the transpose of an orthogonal matrix is the inverse of the matrix,
\be
O^TH_oO=E\,.
\ee
The matrix $H_o$ is thus diagonalized, at the cost of introducing a linear transformation of the random variables:
\be
\frac{K}{N-2s}\sum_j\big(\del2{j}\big)^2=U^TH_oU=U^TOO^TH_oOO^TU=(U^TO)E(O^TU)\,.
\ee
We introduce $N-1$ new random variables by means of the transformation:
\be
V=O^TU\,.
\ee
Then the term in the exponent representing the Gaussian distributions is
\be
-\halb U^T {\bf 1 }U=-\halb U^TO1O^TU=-\halb V^T {\bf 1}V=-\halb \sum_{n=1}^{N-1}V_n^2\,.
\ee
The Gaussian distributions remain basically unchanged.  

Further, the determinant of an orthogonal matrix is $\pm 1$, because the transpose of the matrix is also the inverse:
\be
\det(O^{-1}O)=1=\det(O^TO)=\big(\det(O)\big)^2.
\ee
Therefore, changes in the volume element are simple since the volume element for the new variables is
\ba
dV_1 dV_2 ...dV_{N-1}=\bigg|\det \bigg( \frac{\partial V_m}{\partial U_n}\bigg)\bigg|dU_1dU_2...dU_{N-1}\nonumber\\
=\vert \det(O)\vert dU_1 dU_2 ...dU_{N-1}\nonumber\\ 
=dU_1 dU_2...dU_{N-1}.
\ea
After completing the diagonalization, 
\be
\frac{1}{N-2s}\sum_j\big(\Delta_{j,s}^{(2)}\big)^2=\sum_i \epsilon_i V_i^2\,.
\ee
The probability is therefore
\be
P(A_o)=\int\frac{d\omega}{2\pi}e^{i\omega\big(A_o-\sum_k\epsilon_k V_k^2\big)}\prod_i\bigg( e^{-\frac{V_i^2}{2}}\frac{dV_i}{\sqrt{2\pi}}\bigg)\,.
\ee
An eigenvalue of zero will not contribute in any way to this probability since the random variable corresponding to a zero eigenvalue just integrates out.  

Let the eigenvalue $\epsilon_i$ have multiplicity $\mu_i$, by which is meant that the eigenvalue $\epsilon_i$ is repeated $\mu_i$ times.  Integration over the random variables then gives a useful form for the probability:
\be\label{probint}
P(A_o)=\int_{-\infty}^{+\infty} \frac{d\omega}{2\pi}\frac{e^{i\omega A_o}}{\prod_i(1+2i\epsilon_i \omega)^{\mu_i/2}}\,.
\ee

Finally the contour integral may be deformed and closed in the upper half complex plane where it encloses the singularities of the integrand.  This is discussed in Appendix 1. Knowing the probability, one may integrate with respect to the variance to find the cumulative distribution, and then find the limits corresponding to a 50\% probability of observing the variance. 

	{\it Properties of the eigenvalues\/}.  First, it is easily checked that the probability is correctly normalized by integrating over all $A_o$ and using properties of the delta-function:
\ba
\int P(A_o)dA_o=\int_{-\infty}^{+\infty}
\frac{d \omega}{2\pi} 
\frac{\int e^{i\omega A_o}dA_o}
	{\prod_i(1+2 i \epsilon_i \omega)^{\mu_i/2}}
=\int_{-\infty}^{+\infty}
\frac{\delta(\omega) d\omega }{\prod_i(1+2 i \epsilon_i \omega)^{\mu_i/2}}\nonumber\\
=\int_{-\infty}^{+\infty}d \omega \delta(\omega)=1\,.
\ea
Second, the eigenvalues are all either positive or zero. The eigenvalue equation for the eigenvector labeled by $\epsilon$ is:
\be
\frac{K}{N-2s}\sum_j C^j(C^j)^T\psi_{\epsilon}=\epsilon \psi_{\epsilon}\,.
\ee
Multiply on the left by $\psi_{\epsilon}^T$; assuming the vector has been normalized, we obtain
\be
\epsilon=\frac{K}{N-2s}\sum_j\bigg((C^j)^T\psi_{\epsilon}\bigg)^2\ge 0\,.
\ee
Thus every eigenvalue must be positive or zero.

Next let us calculate the trace of $H_o$.  Since the trace is not changed by an orthogonal transformation,
\ba
{\rm Trace}(O^TH_oO)={\rm Trace}(H_oOO^T)={\rm Trace}(H_oOO^{-1})\nonumber\\
={\rm Trace}(H_o)=\sum_i \epsilon_i\,.
\ea
The sum of the diagonal elements of $H_o$ equals the sum of the eigenvalues of $H_o$.  If we then explicitly evaluate the sum of the diagonal elements of $H_o$ we find
\ba
\sum_i \epsilon_i=\frac{K}{N-2s}\sum_j{\rm Trace}\big(C^j (C^j)^T)\nonumber\\
=\frac{K}{N-2s}\sum_j\sum_{m>0}\bigg((F_m^j\big)^2+\big(G_m^j)^2)\bigg)\nonumber\\
=K\sum_{m>0}
\frac{\bigg(\sin\frac{\pi m s}{N}\bigg)^4}
{\vert f_m \vert^{2-2\alpha}}=\sigma_y^2(\tau)\,.
\ea
Every term labeled by $j$ contributes the same amount.  We obtain the useful result that {\it the overlapping Allan variance is equal to the sum of the eigenvalues of the matrix $H_o$.}  Similar results can be established for many of the other types of variances.

        {\it Distribution of eigenvalues\/}.  The eigenvalue equation $H_o\psi_{\mu}=\epsilon \psi_{\mu}$ produces many zero eigenvalues, especially when $\tau$ is large.  The dimension of the matrix $H_o$ is therefore much larger than necessary.  Numerical calculation indicates that for the completely overlapping Allan variance, the eigenvalue equation has a total of $N-1$ eigenvalues, but only $N-2s$ non-zero eigenvalues; the number of significant eigenvalues is in fact equal to the number of terms in the sum over $j$ in the equations:
\be
\frac{K}{N-2s}\sum_n\sum_j^{N-2s}(C_m)^j(C_n)^j\psi_{n}=\epsilon \psi_m\,.
\ee
The factorized form of $H_o$, that arises on squaring a difference operator,  permits the reduction of the size of the matrix that is to be diagonalized.  We introduce the quantities
\be
\phi_{\mu}^j=\sum_n(C_n)^j\psi_{n\mu}\,.
\ee
We are using the Greek index $\mu$ to label a non-zero eigenvalue and the index $\nu$ to label a zero eigenvalue.  The eigenvalue equation becomes
\be
\frac{K}{N-2s} \sum_j(C_m)^j\phi_{\mu}^j=\epsilon \psi_{m\mu}\,.
\ee
Multiply by $(C_m)^l$ and sum over the frequency index $m$.  Then
\be\label{eigenoverlapJ}
\frac{K}{N-2s}\sum_{m,j}(C_m)^l(C_m)^j\phi_{\mu}^j=\epsilon\phi_{\mu}^l\,.
\ee
This is an eigenvalue equation with reduced dimension $N-2s$ rather than $N-1$, since the number of possible values of $j$ is $N-2s$. The eigenvalue equation can be written in terms of a reduced matrix $H_{red}$, given by
\be
(H_{red})^{lj}=\frac{K}{N-2s}\sum_m(C_m)^l(C_m)^j\,.
\ee
The indices $l,j$ run from 1 to $N-2s$. Eigenvalues generated by Eq. (\ref{eigenoverlapJ}) are all non-zero.  To prove this, multiply Eq. (\ref{eigenoverlapJ}) by $\phi_{\mu}^l$ and sum over $l$.  We obtain
\be
\frac{K}{N-2s}\sum_m\bigg(\sum_l (C_m)^l\phi_{\mu}^l  \bigg)^2=\epsilon\sum_l\big(\phi_{\mu}^l \big)^2\,.
\ee
The eigenvalue cannot be zero unless
\be\label{zerocondition}
\sum_l (C_m)^l\phi_{\mu}^l =0
\ee
for every $m$.  The number of such conditions however is larger than the number $N-2s$ of variables, so the only way this can be satisfied is if $\phi_{\mu}^l=0$, a trivial solution.  Therefore to obtain normalizable eigenvectors from Eq. (\ref{eigenoverlapJ}), the corresponding eigenvalues must all be positive.
        This is true even through some of these conditions may be trivially satisfied if the factor $\sin(\pi m s/N)$ vanishes, which happens sometimes when
\be\label{lossequation}
ms=M N
\ee
where M is an integer.  Every time a solution of Eq. (\ref{lossequation}) occurs, two equations relating components of $\phi_{\mu}^l$ are lost.  Suppose there were $n$ solutions to Eq. (\ref{lossequation}); then the number of conditions lost would be $2n$.  The number of variables is $N-2s$ and the number of conditions left in Eq. (\ref{zerocondition}) would be $N-1-2n$.  The excess of conditions over variables is thus
\be
N-1-2n-(N-2s)=2(s-n)-1\,.
\ee
In Appendix 2 we prove that  under all circumstances $2(s-n)-1>0$.

	We temporarily drop the subscript $o$ since the remainder of the results in this section are valid for any of the variances.  If the eigenvalues are found and the appropriate matrix is diagonalized, we may compute the probability for observing a value of the overlapping variance, denoted by the random variable $A$, by

\ba
P(A)=\int_{-\infty}^{\infty}\frac{d\omega}{2\pi}e^{i\omega\big(A-V^TEV\big)}\prod_i\bigg(\frac{e^{-V_i^2/2}dV_i}{\sqrt{2\pi}}\bigg)\nonumber\\
=\int\frac{d\omega}{2\pi}\frac{e^{i\omega A}}{\prod(1+2i\epsilon_i \omega)^{\mu_i/2}}\,.
\ea
{\it Case of a single eigenvalue\/}.  If a single eigenvalue occurs once only, the general probability expression, Eq. (\ref{probint}), has a single factor in the denominator:
\be
P(A)=\frac{1}{2\pi}\int_{-\infty}^{\infty}\frac{d \omega e^{i\omega A}}{\sqrt{1+2i\omega\epsilon}}\,.
\ee
The integral is performed by closing the contour in the upper half complex $\omega$ plane.  There is a branch point on the imaginary axis at $\omega=i/(2\epsilon)$, and a branch line from that point to infinity.  Evaluation of the integral gives:
\be\label{probmod}
P(A)=\frac{1}{\sqrt{2\pi\sigma_y^2(\tau)}}\frac{e^{-A/(2 \sigma_y^2(\tau))}}{\sqrt{A}}
\ee
This is a chi-squared distribution with exactly one degree of freedom. The computation of the confidence interval for a given $s$ is simple.  The cumulative probability obtained from Eq. (\ref{probmod}) is
\be
\Phi(A)=\int_0^{A}\frac{1}{\sqrt{2\pi\sigma_y^2(\tau)}}\frac{e^{-x/(2 \sigma_y^2(\tau))}}{\sqrt{x}}dx={\rm erf}\big(\frac{A}{\sqrt{2 \sigma_y^2(\tau)}} \big)\,.
\ee
The $\pm 25$\% limits on the probability of observing a value $A$ are then found to be $1.323 \sigma_y^2$ and $0.1015 \sigma_y^2$, respectively. An example of this is plotted in Figure 2.  ($A$ is a variance, not a deviation.)

{\it Case of two distinct non-zero eigenvalues\/}.  For the overlapping variance, when $s$ has its maximum value $N/2-1$ there are two unequal eigenvalues.  The probability integral can be performed by closing the contour in the upper half plane and gives the expression
\be\label{twoeigenvalues}
P(A)=\frac{1}{2\sqrt{\epsilon_1 \epsilon_2}}e^{-\frac{A}{4}\big(\frac{1}{\epsilon_1}+\frac{1}{\epsilon_2}\big)}I_0\bigg(\frac{A}{4}\big(\frac{1}{\epsilon_2}-\frac{1}{\epsilon_1} \big)   \bigg)\,.
\ee
where $I_0$ is the modified Bessel function of order zero.  The probability is correctly normalized.  It differs from a chi-squared distribution in that the density does not have a singularity at $A=0$. This is illustrated in Figure 1. 

Evaluation of the contour integral when there are more than two distinct eigenvalues is discussed in Appendix 1.  If an eigenvalue occurs an even number $2n$ times, the corresponding singularity becomes a pole of order $n$ and a chi-squared probability distribution may result; this has only been observed to occur for white PM.

\section{Variance of the Variance}
	One of the goals of this investigation is to estimate the uncertainty in the variance when some value of the variance is measured.  This can be rigorously defined if the probability $P(A_o)$ is available, but it may be difficult to evaluate the integral in Eq. (\ref{probint}).  In this case, one might be interested in a measure such as the rms value of the variance, measured with respect to the mean variance $\sigma_y^2(\tau)$.  One method for obtaining this measure is obtained from the probability without introducing a Fourier integral representation for the delta function:
\ba
\int A_o^2 P(A_o)dA_o=\int A_o^2\delta(A_o-\sum_i \epsilon_i V_i^2)
\prod_m \bigg(e^{-V_m^2/2}\frac{dV_m}{\sqrt{2\pi}}  \bigg)dA_o \nonumber\\
=\int \bigg(\sum_i \epsilon_i V_i^2 \bigg)^2\delta(A_o-\sum_i \epsilon_i V_i^2)\prod_m \bigg(e^{-V_m^2/2}\frac{dV_m}{\sqrt{2\pi}}  \bigg)dA_o \nonumber\\
=\int \bigg(\sum_i \epsilon_i V_i^2 \bigg)^2
\prod_m \bigg(e^{-V_m^2/2}\frac{dV_m}{\sqrt{2\pi}}  \bigg)\,.
\ea
Expanding the square in the integrand, there are $N$ fourth-order terms and $N(N-1)$ cross terms so we get
\ba
\int A_o^2 P(A_o)dA_o=\sum_i \epsilon_i^2 \big< V^4 \big> + 2 \sum_{i \ne j}\epsilon_i\epsilon_j \big<V^2 \big>^2\nonumber\\
=3\sum_i \epsilon^2+2 \sum_{i \ne j} \epsilon_i \epsilon_j\,.
\ea
To obtain the variance of the variance, we must subtract
\be
\big<A_o \big>^2=\sum_i \epsilon_i^2+2 \sum_{i \ne j} \epsilon_i \epsilon_j\,.
\ee
The result is
\be\label{varofvar}
\big<A_o^2\big>-\big<A_o\big>^2=2\sum_i \epsilon_i^2\,.
\ee
Therefore {\it the rms deviation of the variance from the mean variance equals $\sqrt{2}$ times the square root of the sum of the squared eigenvalues.}  This result is not very useful if there are only a small number of eigenvalues, since the factor $\sqrt{2}$ can make the rms deviation from the mean of the variance larger than the variance.  This is because the probability distribution of the variance is not a normal distribution, and can lead to unreasonably large estimates of confidence intervals.

If the eigenvalues or the probabilities are not readily available, a similar confidence estimate can be obtained from an alternative form of Eq. (\ref{varofvar}) by considering the trace of the square of $H_o$:
\be\label{varofvar2}
2\sum_i\epsilon_i^2=2\sum_{i,j}\big(\sum_{m>0}(F_m^iF_m^j+G_m^iG_m^j) \big)^2\,.
\ee
To prove this, consider diagonalization of $H_o$ by the orthogonal transformation $O$, and compute
\ba
2\sum_i\epsilon_i^2=2{\rm Trace}(E^2)=2{\rm Trace}(O^TH_oOO^TH_oO)=2{\rm Trace}H_o^2\nonumber\\
=\frac{2K^2}{(N-2s)^2}\sum_{m,n}\bigg(\sum_i\big((C_m)^i(C_m)^i\big)\sum_j\big((C_n)^j(C_n)^j\big) \bigg)\nonumber\\
=\frac{2K^2}{(N-2s)^2}\sum_{i,j}\bigg(\sum_m\big((C_m)^i(C_m)^j\big) \sum_n\big((C_n)^j(C_n)^i  \big)  \bigg)\nonumber\\
=\frac{2K^2}{(N-2s)^2}\sum_{i,j}\bigg(\sum_m\big(F_m^iF_m^j+G_m^iG_m^j \big) \bigg)^2\,.
\ea
If there are not too many terms in the sums over $i$ and $j$ then the sum over the frequency index in Eq. (\ref{varofvar2}) becomes
\be
\sum_m\big(F_m^iF_m^j+G_m^iG_m^j \big)=\sum_{m>0}\frac{\big(\sin\frac{\pi m s}{N}\big)^4}{f_m^{2\lambda}}\cos\frac{2\pi m(i-j)}{N}\,.
\ee
This leads to some useful approximation schemes, but these will not be discussed further in this paper.

Usually the variance of the variance is larger than the range of possible values of the variance computed from the $\pm 25$\% limits obtained from calculated probabilities.  An example is given in Figure 1, where the confidence intervals are plotted for the non-overlapping variance for $N=1024$.  For $s$ ranging between 342 and 511 there is only one eigenvalue, so the rms deviation of the variance from the variance is $\sqrt{2}$ times the variance.  This gives an upper limit on the confidence interval that is 1.414 times the variance, larger than that obtained from the actual probability distribution.  The medium heavy lines in Figure 1 show the true $\pm 25$\% probability limits obtained from calculations such as Eq. (\ref{probfunction})
\begin{figure}\label{varsim}
\centering
\includegraphics[width=4.5 truein]{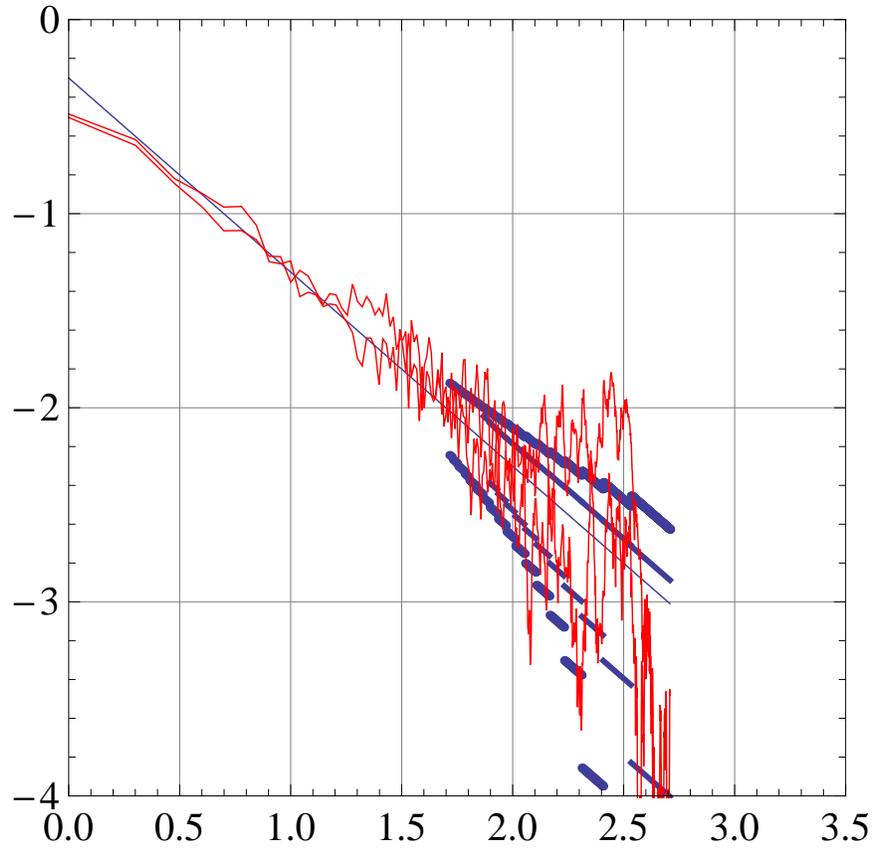}
\caption{Comparisons of average non-overlapping Allan variance (light line) with rms deviations from the mean of the variance (heavy lines) and true $\pm 25$\% limits (medium heavy lines) for $N=1024$ data items in the time series.  Results for the Allan variance for two independent simulation runs are plotted for comparison.}
\end{figure}
\section{Modified Allan Variance}
	The modified Allan variance is defined so that averages over time are performed before squaring and averaging.  We use a subscript $m$ to distinguish this form from the overlapping form of the variance.  The definition is 
\be
\sigma_m^2(\tau)=\bigg<\bigg(\frac{1}{s}\sum_{j=1}^s \del2{j}  \bigg)^2  \bigg>\,.
\ee
Summing the expression for the second-difference given in Eq. (\ref{del2ave}),
\ba
\frac{1}{s}\sum_j^s \del2{j} =-\sqrt{\frac{h_{\alpha}}{2 s^2 \pi^2 \tau^2 \sigma^2(N\tau_0)}}\sum_m\frac{w_m}{\vert f_m\vert^{\lambda}}\big(\sin\frac{\pi m s}{N} \big)^2\nonumber\\
\times\bigg(\frac{\sin\frac{\pi m s}{N}}{\sin \frac{\pi m}{N}} \bigg)e^{-\frac{\pi i m (3s+1)}{N}}\,.
\ea
Squaring this, we write the complex conjugate of one factor and obtain
the ensemble average
\be\
\sigma_m^2(\tau)=\frac{h_{\alpha}}{2 s^2 \pi^2 \tau^2 \sigma^2(N\tau_0)}
\sum_{m,n}\frac{\big<w_mw_n^*\big>}{\vert f_m f_n\vert^{\lambda}}\frac{\bigg(\sin \frac{\pi m s}{N}\bigg)^6}{\big(\sin \frac{\pi m}{N}\big)^2}\,.
\ee
Using Eq. (\ref{wmaverage})and writing the result in terms of a sum over positive frequencies only,
\be\label{modsigma}
\sigma_m^2(\tau)=2\sum_{m>0}\frac{S_y(f_m)}{\pi^2 s^2 \tau^2 f_m^2 (N \tau_0)}
\frac{\bigg(\sin \frac{\pi m s}{N}\bigg)^6}{\bigg(\sin\frac{\pi m}{N}\bigg)^2}\,.
\ee
Passing to an integral when the frequency spacing is sufficiently small,
\be
\sigma_m^2(\tau)=2\int_0^{f_h}\frac{df S_y(f)}{\pi^2 s^2 \tau^2 f^2} 
\frac{\big(\sin( \pi f \tau)\big)^6}{\big(\sin(\pi f\tau_0)\big)^2}\,.
\ee
This agrees with previously derived expressions\cite{sullivanetal}.  The appearance of the sine function in the denominator of this expression makes the explicit evaluation of the integral difficult.  In Table 1 we give the leading contributions to the modified Allan variance for very large values of the averaging time $\tau$. In general there are additional oscillatory contributions with small amplitudes that are customarily neglected.  These results do not agree however with those published in\cite{lesage84}; the leading terms do agree with those published in\cite{howe99}.
\begin{table}
\begin{center}
\begin{tabular}{|c|c|c|c|c|}
\hline
\vbox to .3in{}
Noise Type&$S_y(f)$& Mod $\sigma_y^2(\tau)$&$\lambda$&$\alpha$\\ 
\vbox to -.3in{}&&&&  \\ \hline
\vbox to .0in{}&&&&\\ 
 White PM& $h_2 f^2$ &$\mathstrut$ $  \frac{3h_2}{\pi^2 \tau^3}\big(1 +\frac{5}{18 s^3}\big)$&0&2\\ 
\vbox to .03in{}&&&& \\ \hline
\vbox to .03in{}&&&&\\
Flicker PM& $h_1 f  $&  $\frac{3h_1 \ln(256/27)}{8\pi^2\tau^2}$&$\frac{1}{2}$&1 \\
\vbox to .03in{}&&&& \\ \hline
\vbox to .03in{}&&&&\\
 White FM&$h_0$&  $\frac{h_0}{4\tau}\big(1+\frac{1}{2 s^2}\big)$&1& 0 \\ 
\vbox to .02in{}&&&& \\ \hline
\vbox to .02in{}&&&&\\
 Flicker FM&$  h_{-1}f^{-1}$&$ 2 h_{-1}\ln\big(\frac{3^{27/16}}{4}\big)$&$\frac{3}{2}$& -1 \\ 
\vbox to .02in{}&&&& \\ \hline
\vbox to .02in{}&&&&\\
Random Walk&$h_{-2}f^{-2}$&$h_{-2}\pi^2 \tau\big(\frac{11}{20}+\frac{1}{12 s^2}+\frac{1}{40s^4}  \big)   $&2& -2 \\ 
\vbox to .2in{}&&&&  \\ \hline
\end{tabular}
\caption{Asymptotic expressions for the Modified Allan Variance, in the limit of large sampling times $\tau = s\tau_0$.}
\end{center}
\end{table}

{\it Eigenvalue structure for modified Allan variance\/}.
	For the modified case, instead of Eq. (\ref{hoverlapping}) there are two summations;  we use a subscript $m$ on $H$ to denote the modified form:
\be
H_m=\frac{K}{s^2}\sum_j C^j\sum_k(C^k)^T\,.
\ee
We then seek solutions of the eigenvalue equation
\be
H_m\psi_{\epsilon}=\frac{K}{s^2}\sum_j C^j\sum_k( C^k)^T\psi_{\epsilon}=\epsilon \psi_{\epsilon}\,.
\ee
Multiply by $\sum_l(C^l)^T$ and sum over the frequency index.  
\be
\frac{K}{s^2}\sum_{l,j,m}\big((C_m)^l)^T(C_m)^j\big(\sum_k(C^k)^T\psi_{\epsilon}\big)=\epsilon\big(\sum_l (C^l)^T \psi_{\epsilon}\big)\,.
\ee 
The matrix equation has been reduced to a scalar equation for the quantity
\be
\phi=\sum_k (C^k)^T\psi_{\epsilon}\,.
\ee 
Here the matrix product of $(C^k)^T$ with $\psi_{\epsilon}$ entails a sum over all frequencies.  Since we have a scalar eigenvalue equation for $\epsilon$, there can be one and only one eigenvalue, which is easily seen to be the same as given by Eq. (\ref{modsigma}) for each sampling time $\tau $:
\be
\epsilon=\frac{K}{s^2}\sum_{l,k,m}\big((C_m)^l\big)^T (C_m)^k=\sigma_m^2(\tau)\,.
\ee
The probability distribution will be of the form of Eq. (\ref{probmod}), a chi-squared distribution with one degree of freedom.

\section{Non-overlapping Allan variance}

	In the non-overlapping form of the Allan variance, the only change with respect to Eq. (\ref{overlapvar}) is that the sum over $j$ goes from 1 in steps of $s$ up to $j_{max} \le N-2s$.  We denote the number of values of $j$ by $n_{max}$.  The average non-overlapping variance is the same as that for the overlapping case, but the probability distributions are different. The matrix $H_{no}$ takes the form
\be
(H_{no})_{mn}=\frac{K}{n_{max}}\sum_{j=1,1+2s..}^{j_{max}}(C_m)^j(C_n)^j\,.
\ee
The subscript ``no" labels a non-overlapping variance, and the indices $m$ and $n$ label frequencies.  Here there are only $n_{max}$ terms in the sum since the values of $j$ skip by $s$.  $n_{max}$ is given by:
\be
n_{max}=\lfloor \frac{N-1}{s}\rfloor -1\,,
\ee
where $\lfloor Q \rfloor $ denotes the largest integer less than or equal to $Q$.  In order to diagonalize $H_{no}$ and compute probabilities, we look for eigenvalues by seeking solutions of:
\be\label{eigenno}
\sum_n(H_{no})_{mn}\psi_n=\epsilon \psi_m.
\ee
We define a reduced eigenvector by
\be
\phi_j=\sum_n(C_n)^j\psi_n.
\ee
The eigenvalue equation reduces to
\be
\frac{K}{n_{max}}\sum_j(C_m)^j\phi_j=\epsilon \psi_m.
\ee
Multiply this equation by $(C_m)^l$ and sum over the frequency labels $m$.  Then
\be\label{eigennonoverlap}
\frac{K}{n_{max}}\sum_j\sum_m(\big(C_m)^l(C_m)^j\big) \phi_j=\epsilon \phi_l\,.
\ee
Let
\be
J_{lj}=\frac{K}{n_{max}}\sum_m(C_m)^l (C_m)^j=\frac{K}{n_{max}}(C^l)^T C^j;
\ee
the sum over frequencies is accomplished by the matrix multiplication.
Just as for the overlapping case, if we calculate the trace of the matrix $J_{lj}$,
we find that since each term in the sum over $j$ contributes the same amount,
\be
{\rm Trace}(J_{lj})= \frac{K}{n_{max}}\sum_j C_j^TC_j=\sigma_{no}^2(\tau)\,.
\ee
Because the trace remains unchanged under an orthogonal transformation, the non-overlapping Allan variance will be equal to the sum of the eigenvalues of Eq. (\ref{eigennonoverlap}).  The probability functions for a given $\tau$ can differ from those for the overlapping or modified cases because the number of eigenvalues and their multiplicities may be different.  The eigenvalues are still all greater than or equal to zero.

	If the eigenvalues are found and the matrix $J_{lj}$ is diagonalized, we may compute the probability for observing a value of the overlapping variance, denoted by the random variable $A_{no}$, by
\ba
P(A_{no})=\int_{-\infty}^{\infty}\frac{d\omega}{2\pi}e^{i\omega\big(A_{no}-U^TJU\big)}\prod_i\frac{e^{-U_i^2/2}}{\sqrt{2\pi}}\nonumber\\
=\int\frac{d\omega}{2\pi}\frac{e^{i\omega A_{no}}}{\prod(1+2i\epsilon_i \omega)^{\mu_i/2}}\,.
\ea
For example, when there are only two distinct non-zero eigenvalues, The matrix $J_{lk}$ will have elements
\ba
J_{1,1}=J_{1+s,1+s}=\frac{1}{2} \sigma_{no}^2;\hbox to 1in{}\nonumber\\
J_{1,1+s}=J_{1+s,1}=\int_0^{f_h} \frac{S_y(f) df }{(\pi \tau f)^2} \big(\sin(\pi \tau f)\big)^4\cos(2\pi \tau f)\,.
 \ea
The eigenvalues are then
\ba
\epsilon_1=\frac{1}{2}\sigma_{no}^2+ |J_{1,1+s}|\nonumber\\
\epsilon_2=\frac{1}{2}\sigma_{no}^2- |J_{1,1+s}|\,.
\ea
The probability is of the form of Eq. (\ref{twoeigenvalues}).
\begin{figure}
\label{flickersim}
\centering
\includegraphics[width=5 truein]{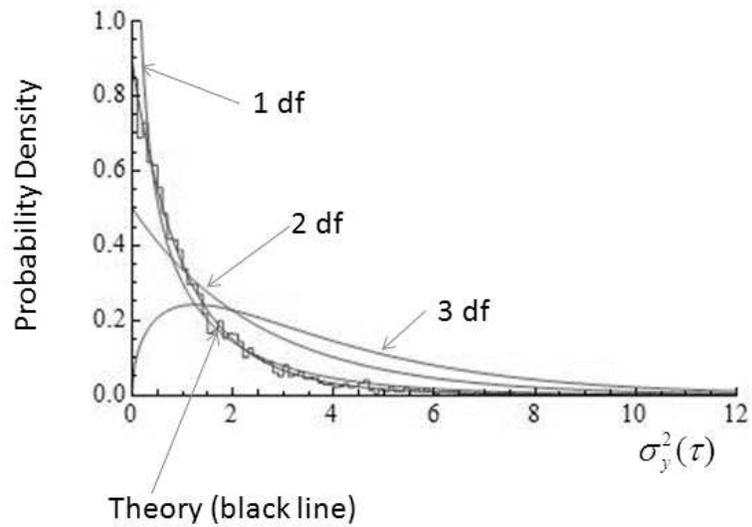}
\caption{Simulation of Allan variance with $N=64$, $s=19$.  For this case there are two distinct eigenvalues in the matrix for the non-overlapping case.  The variance for $s=19$ was extracted from each of 4000 independent runs and a histogram of the values obtained was constructed for comparison with the probability, Eq. (\ref{twoeigenvalues}).  Chi-squared distributions with 1, 2, and 3 degrees of freedom are plotted for comparison.}  
\end{figure}
Figure 1 shows an example of this for flicker FM noise, for $N=64$ items in the time series.  The variance corresponding to $\tau=19 \tau_0$ was extracted from each of 4000 independent simulation runs and a histogram of the resulting values was plotted.  Good agreement with the predicted probability distribution can be seen.

\section{Other variances}

	The analysis methods developed in this paper can be extended to other variances. For example, the Theo variance\cite{Howe03} is defined by an overlapping average,
\ba\label{theo}
\sigma_{Th}^2(s,\tau_0,N)=\frac{1}{N-s}\sum_{i=1}^{N-m}\frac{4}{3\tau^2}\times \hbox to 1.5in{}\nonumber\\
\sum_{\delta=0}^{s/2-1}\frac{1}{s/2-\delta}\bigg((X_i-X_{i-\delta+s/2})- 
	(X_{i+s}-X_{i+\delta+s/2})\bigg)^2\,.
\ea
where $s$ is assumed to be even, and $\tau=3s\tau_0/4$.  For any of the power law noises, and after passing to an integral, this expression can be transformed with the aid of elementary trigonometric identities to
\be
\sigma_{Th}^2(\tau,\tau_0,N)=\frac{8}{3}\int_0^{f_h}\frac{S_y(f)df}{(\pi \tau f)^2}\sum_{\kappa=1}^{2\tau/3\tau_0}\frac{1}{\kappa}\bigg(\sin(\pi f \kappa \tau_0)\sin\big(\pi f (\frac{4}{3}\tau-\kappa \tau_0)\big)\bigg)^2\,.
\ee
The integer $\kappa$ in the denominator makes this variance more difficult to evaluate, but a factorized quadratic form can still be constructed and the eigenvalue structure and resulting probabilities can be analyzed.

The Hadamard variance is defined in terms of a third difference, and is widely used to characterize clock stability when clock drift is a significant issue\cite{hutsell95}.  A third difference operator may be defined as
\be
\Delta_{j,s}^{(3)}=\frac{1}{\sqrt{6\tau^2}}\big(X_{j+3s}-3X_{j+2s}+X_{j+s}-X_j \big)\,.
\ee
The completely overlapping Hadamard variance is the average of the square of this third difference:
\ba
\sigma_H^2(\tau)=\frac{1}{N-3s}\sum_{j=1}^{N-3s}\big(\Delta_{j,s}^{(3)}\big)^2\nonumber\\
\rightarrow\frac{8}{3}\int_0^{f_h}\frac{S_y(f)df}{(\pi \tau f)^2}\big(\sin(\pi \tau f)\big)^6\,.
\ea

	These methods can be applied to cases in which there is dead time between measurements of average frequency during the sampling intervals.  Suppose for example that the measurements consist of intervals of length $\tau=s\tau_0$ during which an average frequency is measured, separated by dead time intervals of length $D-\tau$ during which no measurements are available.  Let the index $j$ label the measurement intervals with $j=1,2,...N$.  An appropriate variance can be defined in terms of the difference between the average frequency in the $j^{th}$ interval and that in the interval labeled by $j+r$: 
\be
\Delta_{j,r,s}^{(2)}=\frac{1}{\sqrt{2}}\big(\overline y_{j+r,s}-\overline y_{j,s}\big)\,,
\ee
where $\overline y_{j,s}$ is the average frequency in the interval $j$ of length $s \tau_0$.
Then an appropriate variance can be defined as
\be
\Psi(\tau,D)=\bigg<\big(\Delta_{j,r,s}^{(2)}  \big)^2  \bigg>\,.
\ee
If the measurements are sufficiently densely spaced that it is possible to pass to an integral, this can be shown to reduce to
\be
\Psi(\tau,D)=\frac{2}{D\tau}\int_0^{f_h}df \frac{S_y(f)}{(\pi f)^2}\big(\sin(\pi f r D)\big)^2 \big( \sin(\pi f \tau)\big)^2\,.
\ee
When $D=\tau$ and $r=1$ there is no real dead time and this variance reduces to the ordinary Allan variance.  

\section{Summary and Conclusion}

	In this paper a method of simulating time series for the common power-law noises has been developed and applied to several variances used to characterize clock stability.  These include overlapping and non-overlapping forms of the Allan variance, and the Modified Allan variance.  Diagonalization of quadratic forms for the average variances leads to expressions for the probabilities of observing particular values of the variance for a given sampling time $\tau=s \tau_0$.  The probabilities are expressed in terms of integrals depending on the eigenvalues of matrices formed from squares of the second differences that are used to define the variance.  Generally speaking, the number of eigenvalues is equal to the number of terms occurring in the sum used to define averages of the second-difference operator, and this number gets smaller as the sampling time $\tau$ gets larger.  The probability distribution $P(A)$ for some variance $A$ is useful in estimating the $\pm25$\% confidence interval about the average variance.  The eigenvalues are usually distinct; only for white PM have eigenvalues been observed to occur with multiplicities other than unity.  This must happen in order that chi-square probability distributions result.  Methods for computing the probabilities have been presented in a few useful cases.  Results presented in this paper are confined to various forms of the Allan variance.

	Other methods of simulating power law noise have been published; the present approach differs from that of\cite{kasdinwalter} in that no causality condition is imposed. The application of the methods developed here will be applied to other variances in future papers. 

\section{Appendix 1. Evaluation of contour integrals for probability\/}

	In almost all cases except for white PM, the eigenvalues are all distinct.  We show here how then the probability function, Eq. (\ref{probint}), can be reduced to a sum of real integrals.  For each of the eigenvalues, we introduce the quantity
\be
r_k=1/(2\epsilon_k)\,.
\ee 
The integral becomes
\be\label{contour2}
P(A)=\prod_k\big(\frac{\sqrt{r_k}}{e^{\pi i/4}} \big)\int\frac{d\omega}{2\pi}\frac{e^{i\omega A}}{\prod_k(\omega-ir_k)^{1/2}}\,.
\ee	
By Jordan's lemma\cite{smirnov}, the contour of integration can be deformed into the upper half plane. The addition of a circular arcs at radius $\vert\omega\vert=\infty$ contributes nothing.  Each of the square root factors in Eq. (\ref{contour2}) has a branch point at $\omega=ir_k$ with a branch line extending to $+i\infty$ along the imaginary axis.  We define the complex argument of each such factor by
\be
-\frac{3\pi}{2} < \arg(\omega-ir_k)<\frac{\pi}{2}\,.
\ee
\begin{figure}
\label{contourplot}
\centering
\includegraphics[width=3. truein]{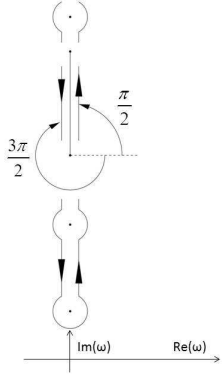}
\caption{Contour deformed to run along branch line on the $y-$axis.  A pair of segments is shown with an odd number of singularities below the segments.  The branch line required by square root factors of the form $\sqrt{\omega-r_k}$ is defined by the angles--either $\pi/2$ or $-3\pi/2$--of the directed segments relative to the Re$(\omega)$ axis.  Contributions from pairs of segments that are above an even number of branch points cancel out.}
\end{figure}

All branch lines extend along the positive $y-$axis to infinity.  The largest eigenvalues give singularities closest to the real axis.  Fig. 3 illustrates the resulting contour.  Around each branch point is a circular portion which contributes nothing because as the radius $\delta$ of each circle approaches zero, the contribution to the integral approaches zero as $\sqrt{\delta}$. The integral then consists of straight segments where $\omega = iy \pm \delta$, where $\delta$ approaches zero.  Two such straight segments are illustrated in Fig. 3, one on each side of the branch line along the $y-$axis.  Suppose the interval of interest is placed so that out of a total of $M$ eigenvalues, $n$ of them are below the interval (three are shown in the figure).  For the contribution to the integral on the left, $y$ is decreasing and we can account for this by reversing the limits and introducing a minus sign.  The phase factor contributed by $n$ factors with branch points below the interval is
\be
\frac{1}{\big(e^{-\frac{3\pi i}{4}}\big)^n}=\bigg(e^{\frac{3\pi i}{4}} \bigg)^n\,.
\ee
The contribution from branch points above the interval of interest is the same on both sides of the $y-$axis, and is
\be
\bigg(e^{\frac{i\pi}{4}}\bigg)^{M-n}\,.
\ee
The factor in front of the integral sign in Eq. (\ref{contour2}) is the same on both sides of the $y-$axis and is
\be
\bigg(e^{-\frac{i\pi}{4}}\bigg)^M\,.
\ee
The total phase factor of this contribution, including a factor $i$ that comes from setting $\omega=iy$ is thus
\be
-i\bigg(e^{-\frac{i\pi}{4}}\bigg)^M\bigg(e^{\frac{i\pi}{4}}\bigg)^{M-n}\bigg(e^{\frac{3\pi i}{4}} \bigg)^n=-i\bigg(e^{\frac{\pi i}{2}} \bigg)^n\,.
\ee
For the integral along the segment on the positive side of the $y-$axis, the only difference is that the phase of each of the $n$ contributions from branch points below the interval changes from $3\pi/4$ to $-\pi/4$.  The phase factor for this part of the contour is thus
\be
+i\bigg(e^{-\frac{i\pi}{4}}\bigg)^M\bigg(e^{\frac{i\pi}{4}}\bigg)^{M-n}\bigg(e^{\frac{-\pi i}{4}} \bigg)^n=+i\bigg(e^{\frac{-\pi i}{2}} \bigg)^n\,.
\ee
If $n$ is even, the two contributions cancel.  If $n=2m+1$ is odd, then the contributions add up with a factor $2(-1)^m$.  The probability is thus always real and consists of contributions with alternating signs, with every other interval left out.

In summary, the contour integral contributions from portions of the imaginary axis in the complex $\omega$ plane that have an even number of branch points below the interval will not contribute to the integral.  For example, if there are four distinct eigenvalues the probability will reduce to 
\ba\label{probfunction}
P(A)=\frac{\sqrt{r_1 r_2 r_3 r_4}}{\pi}\int_{r_1}^{r_2}\frac{e^{-yA}dy}{\sqrt{(y-r_1)(r_2-y)(r_3-y)(r_4-y)}}\nonumber\\
-\frac{\sqrt{r_1 r_2 r_3 r_4}}{\pi}\int_{r_3}^{r_4}\frac{e^{-yA}dy}{\sqrt{(y-r_1)(y-r_2)(y-r_3)(r_4-y)}}\,.
\ea
Such results have been used to evaluate the probabilities for certain sampling intervals for flicker fm noise in Sect. 4.

\section{Appendix 2. Proof that Eq. (\ref{eigenoverlapJ}) generates positive eigenvalues}

In this Appendix we show that under all circumstances Eq. (\ref{lossequation}) gives rise to positive eigenvalues.  

Obviously if $N$ is a prime number Eq. (\ref{lossequation}) can never be satisfied ($n=0$).  Consider the case $s=N-1$.  Then $m=MN$ and there are no solutions since $m \le N/2$.  In general both $m$ and $s$ may contain factors that divide $M$ or $N$.  Suppose $s=ab$ where at least one of the factors $a,b$ is greater than 1, where $a$ divides $M$ and $b$ divides $N$.  Then let
\be
M=am_a,\quad N=bn_b\,.
\ee
Then $m=m_a n_b$. $m=m_a n_b>N/2=bn_b/2$ cannot happen because $m \le N/2$.
If $m_a n_b=N/2=b n_b$, then the number of solutions is $n=1$ and
$2(s-n)-1=2(ab-1)-1 \ge 2(2-1)-1>0$.  If $m=m_a n_b < N/2=b n_b/2$, then there may be solutions $m=m_a n_b, m=2 m_a n_b, m=3m_a n_b,...$up to $b n_b/2$.  The number of such solutions is
\be
n=\lfloor \frac{bn_b}{m_a n_b}\rfloor=\lfloor \frac{b}{m_a}\rfloor
\ee
where $\lfloor x \rfloor$ means the largest integer less than or equal to $x$.  The excess of conditions is then
\be
2(s-n)-1=2(a b -\lfloor \frac{b}{m_a} \rfloor) > 1\,,
\ee
which proves the assertion.

\end{document}